\documentclass[11pt,article]{revtex4-1}
\usepackage{xcolor}
\usepackage{graphicx}
\usepackage{dcolumn}
\usepackage{bm}
\usepackage{comment}
\usepackage{hyperref}
\usepackage{tikz}
\usepackage{changepage}
\usepackage{braket}
\begin{document}

\title{Planet Formation} 

\author{Ravit Helled}
\affiliation{Institute for Computational Science, Center for Theoretical Astrophysics \& Cosmology\\
University of Zurich, Winterthurerstr. 190
CH-8057 Zurich
Switzerland}
\author{Alessandro Morbidelli}
\affiliation{Observatoire de la Cote d'Azur, 
CS 34229\\
06304 Nice Cedex 4, France}

\maketitle
\nopagebreak

\section{Introduction} 
Our galaxy is full with planets. We now know that planets and planetary systems are diverse and come with different sizes, masses and compositions, as well as various orbital architectures. 
Although there has been great progress in understanding planet formation in the last couple of decades, both observationally and theoretically, several fundamental questions remain unsolved. 
This might not be surprising given the complexity of the process that includes various physical and chemical processes,  and spans huge ranges of length-scales, masses, and timescales.  In addition,  planet formation cannot be directly observed  but has to be inferred by gluing together different pieces of information into one consistent picture. 
Observations can provide information on the small solids \citep{2020arXiv200105007A} in young protoplanetary disks or information on mature planets, but the actual  processes that determine how solids and gas accumulate to form planets remains obscure  to measurements. The first planets embedded in protoplanetary disks are just being discovered \cite[e.g.,][]{2018A&A...617A..44K}.
\par 
Planets are formed in protoplanetary disks around young stars \cite{2011ARA&A..49...67W,2014prpl.conf..339T,2014prpl.conf..475A}. 
Such disks are also diverse in terms of mass, size, and lifetime, which from the beginning may produce some diversity in the final planetary systems. 
It is the disk's composition and structure that determines the initial conditions for planetary formation and, once objects are formed in the disk, the interaction with the disk is of significant importance. 
Understanding the physical and chemical properties of protoplanetary disks is crucial since they determine the amount of different solid materials at given radial distance and time, and their lifetimes provide upper limits for the available timescale for giant planet formation.

\section{State of the art}  

{\it How do planets form?} remains a fundamental question in modern astrophysics.  
Studies on the origin of planets have been conducted throughout the last few decades, although thoughts about the formation of the solar system started much longer ago, with the nebular hypothesis being introduced by Immanuel Kant (1755) and Pierre-Simon de Laplace (1796).   
A substantial milestone was the publication of Victor S.~Safronov in 1969 \cite{1969edo..book.....S} where a physical model for planet formation has been presented.  

Until the mid-90s, before the first exoplanet around a sun-like star had been discovered \citep{mayor_queloz_1995}, all planet formation models aimed to explain the properties of the solar system, with the implicit underlying assumption that every planetary system should have had a structure similar to ours. 
Even today, the solar system remains a key benchmark for planet formation models, because we have a huge amount of information on its planets and its small bodies that allow testing the models in  great detail.  

The solar system, however, is a rather well-arranged planetary system with four terrestrial planets, two gas giants and two ice giants, in order of distance from the Sun. 
This leads to a mass function that has a maximum at Jupiter's location.  
The orbits of the planets have small inclinations relative to the  plane orthogonal to the total angular momentum of the system, which itself is almost aligned with the Sun's equator. In addition, the orbits of all planets are almost circular and most of the planets rotate around their spin axis in the same direction they orbit the sun. The farthest planet, Neptune is located at 30 AU and the closest, Mercury, at $\sim 0.4$~AU. Formation models were designed to explain these features as a generic and necessary outcome of the planet-formation process. 

Our view of planetary systems completely changed with the discovery of exoplanets. New types of planets have been found - super-Earths and mini-Neptunes but also super-Jupiters; moreover, giant planets have been found on orbits very close to the host stars (hot Jupiters) but also extremely far from the star at tens of AUs, often on very eccentric orbits very inclined relative to the stellar equator; many packed systems of intermediate-mass planets have been discovered, confined within the orbit of Mercury, often in or near mean-motion resonances with each other. In summary, planetary systems can be dramatically different from our own. 
These observations challenge planet formation models tuned to explain the architecture of the Solar System. Modern planet formation models need to explain not only the detailed structure of the Solar System but also the wide  diversity among planetary systems, as well as the trends observed in terms of stellar type, metallicity, etc.

The standard model for planet formation is known as ``core accretion" \cite{1982P&SS...30..755S,Bodenheimer+86,Pollack1996,Alibert2005, 2009Icar..199..338L,  2010exop.book..319D,2000prpl.conf.1081W,2018haex.bookE.140D,Helled2014, 2014prpl.conf..691B}. 
The first stage of planetary formation is the build-up of a planetary ``core" made of heavy-elements. The growth of the core could be a result of accretion of planetesimals \cite{Pollack1996}, pebbles \cite{Lambrechts2012} or both \cite{Alibert2018}. 
This formation stage leads to the formation of solid proto-planets, often called {\it planetary embryos} if they are sub-terrestrial in mass or {\it planetary cores} if the are more massive. Their mass depends on local conditions (planetesimal density, pebble flux, etc.). 
The growth of planetary embryos/cores is called {\it phase-1}. 
It should be noted that during phase-1, if the mass of the forming object exceeds $\sim$ 2 M$_{\oplus}$ accretion of H-He can take place \cite{valleta18,Brouwers20}. Nevertheless, in this stage the composition of the object is clearly dominated by heavy elements, where $M_z \gg M_{H-He}$. 
If the growth of the core is rapid enough, and the gaseous disk is still present, the forming object can start to steadily accrete gas from the disk ({\it phase-2}). During this stage, mini-Neptunes and Neptune-like planets can be formed, i.e., planets that are still heavy-element dominated but with bound envelopes of H-He. {At the end of this phase the H-He mass can be comparable to the heavy element mass, i.e., $M_z \sim M_{H-He}$.} 
Once the heavy-element mass and the mass of the
gaseous envelope become comparable (known as crossover mass), the growing planet begins to accrete H-He gas in an exponential fashion ({\it phase-3}, a.k.a. {\it runaway}). Planets that reach this formation stage rapidly become giant planets, i.e., planets dominated in mass by the gaseous envelopes {where $M_z \ll M_{H-He}$}, like Saturn and Jupiter. 
Runaway accretion of gas terminates either by dissipation of nebular gas or by other mechanisms that regulate the gas accretion as we discuss below.  
Once the giant planet reaches its final mass it contracts and cools down on a time scale of $10^9$ years (long-term planetary evolution).   
Figure 1 shows sketches of the three main phases of planet formation in the core accretion scenario and the expected formed objects at each stage.
\par

\begin{figure}[h!]
\begin{center}
\includegraphics[width=0.65\textwidth]{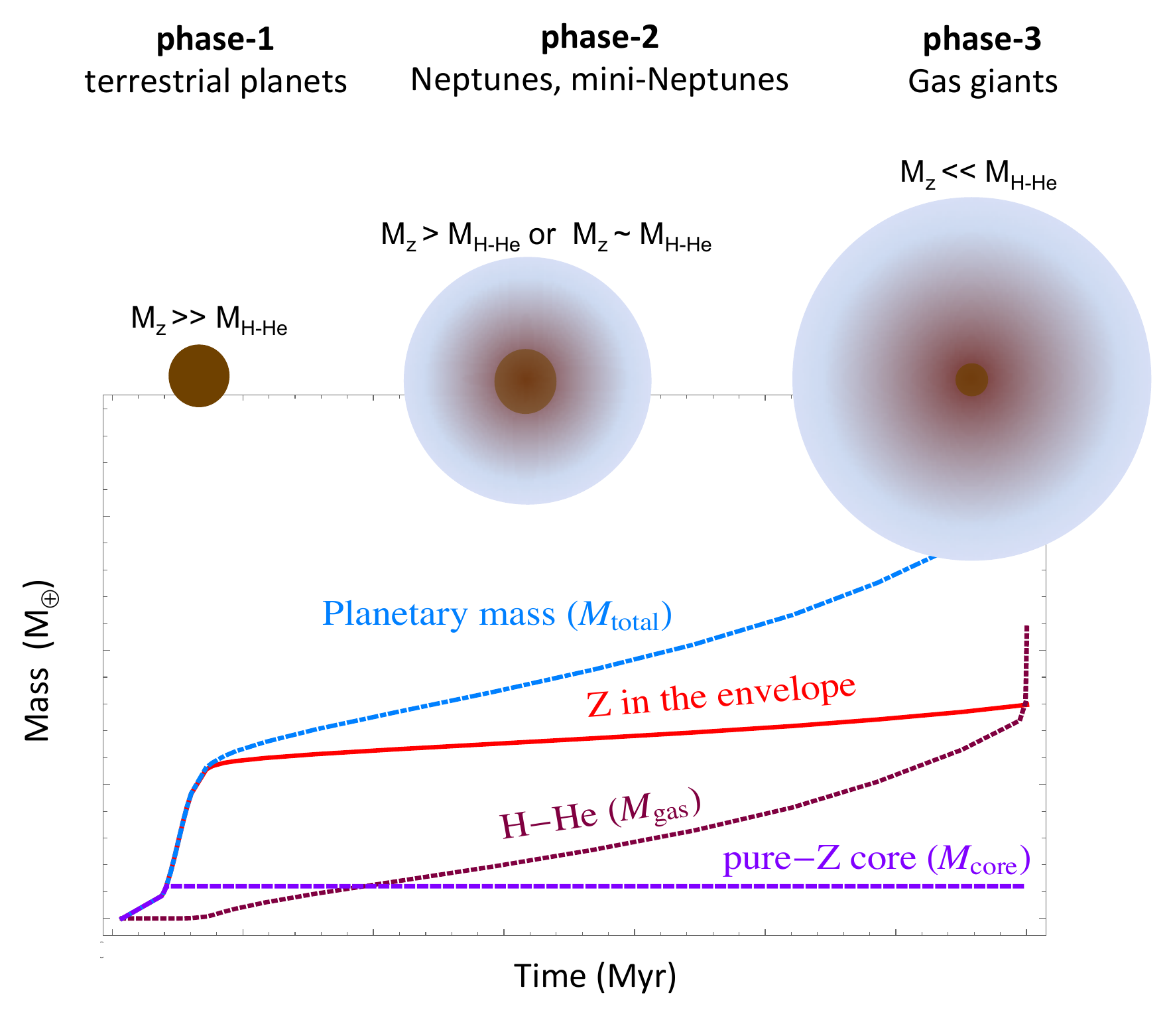}
\end{center}
\caption{\small A sketch of the growth of a  planet  in the core accretion model.
Shown is the planet's mass  vs.~time up to the onset of {\it phase-3}, when runaway gas accretion begins.  
 \textit{Dashed purple line}: pure heavy-element actual core mass. 
\textit{Dashed brown line}: gaseous (hydrogen and helium only) mass. \textit{Solid red line}: heavy-element mass in the envelope.  \textit{Dashed blue line}: total planetary mass.
Sketches of the forming object at each stage  are shown above (not to scale). M$_z$ and M$_{H-He}$ correspond to the heavy-element and H-He  mass, respectively.}
\end{figure}

\clearpage
\section{Important questions} 
Below we list some of the key open questions in planet formation theory.  

\subsubsection*{Where do planets form?} 
It was assumed for decades that the solar system planets formed where we observe them today. The discovery of hot Jupiters and later of smaller planets at very short periods suggested that planets can (significantly) change their orbital location. 
The most accepted explanation is orbital migration when the gaseous disk is still present. In fact planet migration in a disk was already predicted in 1980  \cite{1980ApJ...241..425G}, but it attracted the attention of the community only after the discovery of hot Jupiters  \cite{2002ApJ...565.1257T}. 
Other scenarios that can explain the existence of hot Jupiters are mutual scattering of a system of giant planets \cite{2012ApJ...751..119B} or {\it in situ} formation \cite{2016ApJ...829..114B}. 
Nevertheless, the existence of systems of medium-mass planets in resonance with each other (e.g. Trappist-1, Kepler-223) is a clear signature of migration. Also the heavy-element enrichment of warm Jupiters \cite{Shibata+19b} supports the scenario of planetary migration. 

From the formation point of view, the ideal conditions for the formation of giant planets are near the ice line. However, when migration is included, giant planets forming at the snowline should all end up as hot Jupiters. This is in conflict with the observations suggesting that hot Jupiters are only $\sim$10\% of that of cold Jupiters (i.e. Jovian-mass planets beyond 1 AU from the central star), once observational biases are considered {\bf \cite{2008PASP..120..531C}\cite{2011arXiv1109.2497M}}. 
With the currently expected migration rates, cold Jupiters should have started to form beyond $\sim 20$~AU \cite{2015A&A...582A.112B} in order to migrate no closer to the star than their current orbits within the lifetime of the protoplanetary disk. 
However, it is still unclear where the planets that should have started to form at the snowline might have ended up. Therefore, it is fair to say that this topic is still being debated, and the link between planetary composition and formation location is being investigated intensively. 

Interestingly, although the solar system is a well arranged system, some of its observed properties can be explained if the planets moved around. The orbital structures of the populations of small bodies in the outer solar system require to have been sculpted during a phase of dynamical instability of the giant planets (see \cite{2018ARA&A..56..137N} for a review). This dynamical instability is thought to have occurred after the disappearance of the gas from the protoplanetary disk and increased the mutual separations of the planets' orbits, also exciting somewhat their eccentricities and inclinations. During the gas disk phase, it has been proposed that Jupiter migrated down to 1.5-2 AU before migrating outwards due to the effects of Saturn, as a means to explain the small mass of Mars and the depletion of the asteroid belt \cite{2011Natur.475..206W}. {Even if the migration of Jupiter might not have been so extreme, Saturn is expected to have played a role in slowing down or reversing Jupiter's migration \cite{2001MNRAS.320L..55M} and that all the outer planets entered in mean motion resonance with each other \cite{2007AJ....134.1790M}.  
Unfortunately, there is no direct evidence that constrains the dynamical evolution of the young solar system. The uncertainties on planet migration challenge our ability to link the currently observed planetary orbits to their original formation locations and to assess the importance of dynamical interaction in young planetary systems.  
\subsubsection*{What is the dominating size of solids that build-up planetary embryos/cores?}
It is yet to be determined whether the formation of an embryo/core is dominated by the accretion of large (planetesimals) or small (pebbles) objects, or both.  
The growth of the core via planetesimals 
should stall once all planetesimals in the planet's vicinity have been accreted or scattered away \cite{2010AJ....139.1297L} although when planets form in systems, scattering of planetesimals by one forming planet can brings planetesimals to the other forming planet which is forming nearby \cite{1993Icar..106..210I}. 
Giant planet formation models with planetesimal accretion suggest that the formation of giant planets takes a few Myr. However, there is a large uncertainty on this estimate since the planetary  formation timescale can be decreased substantially if opacity reduction \cite{2010Icar..209..616M} and/or envelope enrichment \cite{2016A&A...596A..90V} are considered. 
Unlike planetesimals, pebbles, being small objects, experience fast orbital decay due to gas drag and therefore continuously resupply solid material to the planet's vicinity; the core's accretional cross-section is also larger for pebbles than for planetesimals \cite{Lambrechts2012}. For pebble accretion the core formation timescale is inversely related to the flux of pebbles, which is unknown and depends on the dust growth rate and the disk's properties (e.g., size, density). Moreover, also the dynamical excitation of the growing planets affects the pebble accretion rate. 
In some cases formation via pebble accretion can be so efficient, that it overestimates the relatively low frequency of giant planets.  Mechanisms that can delay/prevent substantial gas accretion are still being investigated.

A combination of the two scenarios, core formation via pebble accretion followed by  accretion of planetesimals, can explain the growth history of Jupiter deduced from meteoritic constraints (i.e. the rapid growth to $\sim$ 20 M$_{\oplus}$  within 1 My, followed by a slow growth reaching $\sim$ 50 M$_{\oplus}$ not before 3 My) \cite{2017PNAS..114.6712K}.  This was investigated in detail by \cite{Alibert2018} and \cite{2020A&A...634A..31V}  where it was shown that these constraints can be fulfilled only if a heavy-element core is rapidly formed by pebble accretion, followed by a protracted phase of planetesimal accretion, at a rate of $\sim 10^{-6} - 10^{-5}$  M$_{\oplus}$/yr, which delays the onset of runaway gas accretion for $\sim$3 Myr. In this scenario Jupiter's core mass is between 6 and 15 M$_{\oplus}$ with a heavy-element total mass of 20-40 M$_{\oplus}$. 
This formation scenario can also help explaining Jupiter's fuzzy core. 

In the context of giant planet formation, whether dominated by pebble or planetesimal accretion, the forming core is expected to be composed of mostly heavy elements but can also have small amounts of H-He as we discuss above. 
Population synthesis models provide different predictions when assuming growth via pebbles or planetesimals \cite{2020arXiv200604121B}. Clearly a more detailed investigation of the topic is required. We encourage studies that combine both types of solids and investigate the transitions between the two and their dependence on the formation environment.

\subsubsection*{How do planetesimals form?}
Although it is still unknown whether the planetary growth is dominated by planetesimal accretion as discussed above, it is clear that planetesimals must form in protoplanetary disks. We see evidence for this in the Solar System, where asteroids and Kuiper belt objects are the leftover of planetesimal populations that were presumably much larger originally. In extrasolar systems, indirect evidence for massive planetesimal populations comes from the observations of debris disks. In these disks the dust must be continuously produced and mutual collisions among planetesimals is the favored mechanism for dust generation \cite{2014prpl.conf..521M}.  Nevertheless, the formation mechanism for planetesimals is still being actively investigated. 

Very small grains can grow by collisional coagulation, sticking to each other by electrostatic forces. However, when the grains reach a size of about a millimeter,  bouncing becomes the dominant outcome of collisions and further growth is inhibited \cite{2010A&A...513A..57Z}; this is known as ``the bouncing barrier".  The exact size of the bouncing barrier depends on the grains' compositions (see \cite{2010A&A...513A..56G} 
and references therein for details). 
In addition, particles that are large enough to partially decouple from the motion of the gas experience rapid inward drift due to gas drag. This in turn leads to large relative velocities between particles with different sizes. As a consequence collisions become disruptive. 
Therefore collisions among dust particles are unlikely to lead to the formation of planetesimals, i.e., of objects of several kilometers in size.
Moreover, even if somehow disruptive collisions could be avoided, the particles would rapidly drift into the host star. The drift speed is maximal for meter-size boulders, so this problem is known as the ``meter-size barrier".

Great advances on planetesimal formation have been made in the last 15 years, as it was discovered that dust can form clumps due to a number of hydrodynamical effects \cite[e.g.,][]{2005ApJ...620..459Y,2006ApJ...643.1219J,2007Natur.448.1022J,2008ApJ...687.1432C, 2020ApJ...892..120H}. The most promising and deeply studied of these effects is called  the ``streaming instability". 
Regardless of the actual dust-clumping mechanism, when a clumps becomes dense enough the dust remains bound by self-gravity against diffusion generated by turbulence and eventually settles to form a macroscopic object, with a characteristic size of $\sim$~100 km.  
The problem is that efficient dust clumping can be triggered only if (i) the dust particles are large enough (decimeter-size, or Stokes' number of order unity) or (ii) the dust/gas ratio is a few times that of the average protosolar material \cite[e.g.,][]{2009ApJ...704L..75J,2017A&A...606A..80Y,2020ApJ...892..120H}.

Therefore, planetesimal formation by this mechanism is possible only where there has been some preliminary pile-up of dust in the disk. Models of dust evolution show one location where sufficient pile-up should occur naturally: the snowline \cite[e.g.,][]{2016A&A...596L...3I,2017A&A...602A..21S,2017A&A...608A..92D}. But, the planetesimals of the Solar System are so different from each other in terms of chemical and isotopic compositions that they all could not have formed at the same location. In addition, there are still no models capable to predict the resulting radial mass distribution of planetesimals. Therefore, it is fair to say that we do not yet understand how planetesimals form throughout (i.e., not just at one or a few distinct locations) the Solar System and exoplanetary systems. 

 \subsubsection*{What is the role of gravitational instability in planet formation?}
An alternative model for giant planets formation is ``disk instability", where gaseous planets are formed by a local gravitational instability in the protoplanetary disk \cite[see e.g.,]{Boss1997,Mayer2002,Durisen2007,2009ApJ...695L..53B}. This model seems to be required to explain the formation of massive gaseous planets (above a few Jupiter masses), the existence of giant planets at very large radial distances such as in the HR8799 planetary system \citep[e.g.,][]{2010Natur.468.1080M}, the formation of planets in very short timescales \citep[e.g.,][]{2010Sci...329...57L}, and the existence of giant planets around M stars \cite[e.g.,][]{2019Sci...365.1441M}. 

The ``disk instability" model is most efficient in massive disks and in cold regions (i.e. at large radial distances), as implied by the Toomre criterion \cite{1964ApJ...139.1217T} that is used to infer the conditions for disk fragmentation (see \citep{Helled2014} for review).  
However, the efficiency of the ``disk instability" model, and the resulting planetary masses and their compositions, as well as the planetary orbital properties are still being investigated \cite{Nayakshin2017,2018MNRAS.474.5036F,2018ApJ...854..112M}. 

Traditionally, the existence of a heavy-element core and non-stellar compositions of giant planets was used to discriminate between ``core accretion" and ``disk instability". 
However, it was shown that a core can form in both models and that the expected enrichment of giant planets is strongly affected by the formation environment and the early evolution of the planets (e.g., migration followed by planetesimal accretion, atmospheric loss) in both formation models. Therefore, the planetary composition and internal structure cannot be used in a simple manner to discriminate among these models. 

A key observational constraint for giant planet formation models is the correlation between stellar metallicity and the occurrence rate of giant planets \citep{2005ApJ...622.1102F}. 
This correlation can easily be explained by the ``core accretion" model as more metal-rich disks lead to more efficient core formation followed by gas accretion. For disk instability, it is yet to be determined whether disk fragmentation is more likely in metal-rich disks \cite[see][for details]{Helled2014}. 
Nevertheless, it should be noted that this strong metallicity correlation does not exist for massive giant planets that are several Jupiter-mass \citep[e.g.,][]{2017A&A...603A..30S}. 
This may imply that massive giant planets are formed by ``disk instability" while Jupiter-mass planets are formed  via ``core accretion". 

At the moment the main challenge of the ``disk instability" model is to demonstrate that planetary-mass objects can form (and not accrete much during their early evolution) and survive in realistic disk conditions  \citep{2016ARA&A..54..271K}. This is still work in progress and, because the results strongly depend on the numerical model, its resolution and assumptions, tendencies and correlations in the ``disk instability" model are still  unclear.  
However, even if ``disk instability" is not the main mechanism for giant planet formation, gravitational instabilities in young disks may represent the first trials of planet formation, affecting the disk's evolution and the conditions for effective planet formation at later times. 

\subsubsection*{What are the expected compositions and structures  of giant planets?} 
The compositions of giant planets can range from super-solar to solar and even  sub-solar.  
The mass-radius relations of Jupiter and Saturn hint that they contain heavy elements in larger proportion than the protosolar gas; moreover the enrichment in heavy elements is different for the two planets. 
The diversity of planetary compositions of giant exoplanets can be probed for non-highly irradiated planets with measured masses and radii \cite{2011ApJ...736L..29M,2019RNAAS...3..128T}. 
Many of these planets have heavy-element mass fractions that are much higher than the ones predicted from standard formation models. 
During the formation stages 1 and 2, the heavy-element mass is expected to be of the order of 10 M$_{\oplus}$, and accretion of heavy elements during runaway gas accretion should be very inefficient \cite{Shibata+19a}. Therefore, unless the solar nebula was extremely metal-rich \cite{2020A&A...634A..31V} or giant impacts are invoked \cite{2015MNRAS.446.1685L}, it is not easy to explain these enrichments. 
For giant exoplanets an appealing explanation is planetesimal accretion during inward migration \cite{Shibata+19b}. 
Since exoplanet observations are still biased towards close-in planets, the metallicities of giant planets on wider-orbit is yet to be determined. Such measurements would allow us to understand the dependence of the enrichment on the orbital distance.  

The distribution of the heavy elements in giant planet interiors is still being investigated \cite{2018haex.bookE..44H}. 
While the assumption of a simple core+envelope structure is convenient from the modeling perspective, updated formation and structure models of giant planets show that the heavy elements are not uniformly mixed \cite{2012A&A...540A..20L,2017ApJ...836..227L}.
In addition, giant planets probably do not have distinct compact cores; they rather have an innermost region which is highly enriched with heavy-elements. 
This is because once the core mass becomes sufficiently high, the protoplanet can accrete H-He (phase-2) and the subsequently incoming solids dissolve in the atmosphere \cite{HelledStevenson17,valleta18,Brouwers20}. 
This leads to  composition gradients in the deep interiors. Indeed, a fuzzy core seems to exist in Jupiter, as suggested by detailed structure models that fit its gravity field \cite{wahl2017,Debras19}.  
In order to connect the formation and current-state structure it is important to model the long-term planetary evolution and investigate whether/how the heavy-element distribution changes with time \cite{2018A&A...610L..14V,2020arXiv200413534M}. 

\subsubsection*{What controls the final mass of giant planets?} 
Giant planets come at different masses and it is still not understood what mechanism stops the gas accretion. Originally, it was assumed that the planet stops accreting gas when the gaseous disk disappears. 
Giant planets open gaps in the disk, but gap opening is not an effective way to shut-off gas accretion, because the gas penetrates into the gap from the surface of the disk \cite{2014Icar..232..266M}. With or without the gap, the flux of gas to the planet's orbit is essentially the same as the accretion rate of the star from the disk. With a typical value of $10^{-8} M_\odot$/yr, a planet like Jupiter can grow in $10^5$~yr \cite{1999ApJ...526.1001L,2003ApJ...586..540D, 2007ApJ...667..557T, 2009Icar..199..338L, 2016ApJ...823...48T, 2019A&A...630A..82L, 2020ApJ...891..143T}. Thus, there is the need to understand how most giant planets limited their growth to a mass comparable to Jupiter's or just a few times larger. {In the Solar System, we still need to understand how Jupiter and Saturn reached their terminal masses.}

\subsubsection*{What is the role of giant impacts in planetary formation?}
In the solar system it seems that giant impacts played a key role in shaping the planetary properties.  Giant impacts are favorable explanations for the formation of the Earth's moon \cite{2001Natur.412..708C}, the high iron-to-rock ratio in Mercury \cite{2018ApJ...865...35C}, Uranus' obliquity and regular satellites \cite{2018ApJ...861...52K,2020MNRAS.492.5336R,2020arXiv200313582I}, and even Jupiter's fuzzy core \cite{Shang19}, Mars hemisphere dichotomy \cite{2008Natur.453.1216M} and its moons \cite{2018SciA....4.6887C}, and the Pluto-Charon system \cite{2005Sci...307..546C}. This suggests that too many protoplanets tend to form in the disk, so that the system is initially unstable. The impacts reduce the number of massive objects and increase the masses of the survivors, until a stable configuration is obtained. 

Therefore it is likely that giant impacts also played an important role in exoplanetary systems \cite{2015MNRAS.448.1751I,2019NatAs...3..416B}. The fact that the planetary properties can be changed due to giant impacts suggest that one has to be careful when linking the currently observed properties with the planet's origin. 

\subsubsection*{How long does it take for planets to form?} 
Giant planets are H-He dominated and therefore must have completed their formation within the lifetime of the protoplanetary disk (because they accreted the gas from the disk). Since the average age of protoplanetary disks is $\sim$ 3 Myr \cite{2009AIPC.1158....3M}, giant planets are expected to form within this timescale. However, in terms of formation models a couple of Myr make a great difference in the required disk's conditions (which determine the accretion rates), the formation efficiency, the early evolution, as well as the predicted masses and compositions of giant planets in various formation models  \cite{Pollack1996,Alibert2005,2010Icar..209..616M,Helled2014,2014ApJ...789...69H,  Alibert2018,2018SSRv..214...38P,2018haex.bookE.143M,2019A&A...623A..88B,2020A&A...634A..31V}.

The terrestrial planets might have formed on a longer timescale than the disk's lifetime. The Moon-forming event happened several 10 My after gas photo-evaporation, according to radioactive chronometers \cite{2009GeCoA..73.5150K}. 
Mars, instead, ended its accretion within 4-5 My \cite{2011Natur.473..489D}. However, it is not known what was the mass of the Earth when the gas disappeared. In the classic model of Wetherill \cite{1992Icar..100..307W}, the Earth should have had a mass comparable to Mars, but modern models, accounting also for pebble accretion, argue for a much more substantial mass \cite{2018Natur.555..507S,2018MNRAS.479.5136P,2020Icar..33913551L}. The Earth must have been sufficiently massive to retain the material delivered by the cometary shower associated to the giant planet instability, whose amount is constrained by comparing the elemental and isotopic compositions of noble gases in the Earth’s atmosphere with those of comet 67P/C-G \cite{2016E&PSL.441...91M,2017Sci...356.1069M}.

\subsubsection*{How do intermediate mass/size planets form?} 
Exoplanet data suggest that planets with intermediate masses (a few to a few tens of M$_{\oplus}$) and radii (1.5-4 R$_{\oplus}$) are extremely  common in the galaxy. These planets are often refereed to as super-Earths and mini-Neptunes. The first, represent planets with relatively high densities (refractory materials), some kind of large-scale Earth, while the latter represent planets with non-negligible amounts of H-He atmospheres, small or comparable version of Uranus/Neptune.  It is not easy to learn about the formation of such planets without having tighter constraints on their bulk compositions and their original formation location. 
With accurate measurements of the planetary mass and radius, the average planetary density can be determined, and can be used to (non-uniquely) infer the planetary bulk composition. 

The formation of intermediate mass/size planets remains a challenge to planet formation models \cite{2014ApJ...789...69H}. 
The formation of mini-Neptunes and super-Earths is expected to occur at larger radial distances (via pebble/planetesimal accretion) followed by inward migration \cite{2015A&A...582A.112B,2017ApJ...848...95V}. 
If intermediate-mass planets form at large radial distances and migrate inward, they are expected to contain a large amount of volatile elements, water in particular, because their formation started at the snowline. 
However, the analysis of the M-R relationships for super-Earths that are up to $\sim$ 3 M$_{\oplus}$ and radii smaller than 2 R$_{\oplus}$and have no atmosphere indicate that these planets are made of refractory elements \cite{2017ApJ...847...29O,2018ApJ...853..163J,2018ApJ...866...49L}. This could imply that these planets have formed {\it in-situ} from the accumulation of refractory dust at the inner edge of the protoplanetary disk \cite{2014ApJ...780...53C,2019A&A...630A.147F}. 

Future measurements of atmospheric compositions could assist in this direction as well as advances in theory. Because intermediate mass/size planets lie between terrestrial and giant planets in a mass-radius diagram \cite{2020A&A...634A..43O}, knowledge of both groups is required and we suggest that this is a topic that should be investigated in detail in future research, as the return is expected to be enormous. 

\subsubsection*{Have we discovered extrasolar terrestrial planets yet?}
Because of observational biases, planets as small as the terrestrial planets of the Solar System have been discovered only around low-mass stars. 
Around solar-mass stars, so far only super-Earths have been detected, where  some of them are expected to have Earth-like compositions. 
Due to observational biases, in all cases the discovered planets are significantly closer to the parent star than our own terrestrial planets, although some of those orbiting low-mass stars are also in the so-called habitable zone. However, in terms of formation process, it is unclear whether  these planets had formation paths similar to the the terrestrial planets of the solar system \cite{2019A&A...627A..83L}. 
If we identify``terrestrial planets" by their formation process the question in the title of this section has all its relevance. It also highlights the need to identify which of the specific aspects of Earth's formation have been fundamental to determine the properties that make our planet a truly habitable world and to assess the likelihood that the same happened on other planets. 

\section{Opportunities} 
The near and far future seem bright. The ongoing and future observations of planets and planetary systems, as well as the theoretical developments, are expected to significantly improve our understanding of planet formation.  
Some of the key opportunities are listed below:
\begin{itemize}
\item[+] The ongoing TESS (NASA) and CHEOPS (ESA) \cite{2013EPJWC..4703005B} missions as well as the availability of new spectrographs for improved radial velocity measurements (e.g. ESPRESSO, ESO), and coronographs for direct imaging (e.g., SPHERE+, ESO). All these instruments will broaden the census of extrasolar planets and extend it to a region of the mass-distance diagram not yet sampled.  
\item[+] The PLATO ESA mission \cite{2014ExA....38..249R} will provide an understanding of the architecture of planetary systems, probing a larger range of radial distances, with well constrained ages of the systems. 
This can be used to further constrain planet formation and evolution models.  
\item[+] ALMA \cite{2016ApJ...828...46A} keeps providing data on protoplanetary disks, the birth environments of planets. This is then used to constrain the formation timescale and the expected properties of protoplanetary disks and their diversity. 
\item[+]  JWST \cite{2016ApJ...817...17G} and ARIEL \cite{2018ExA....46..135T} are expected to provide key information on the atmospheric composition of gaseous-rich planets. This, together with an accurate determination of the planetary mean density can further constrain the planetary bulk composition, as well as their formation and evolution histories. 
\item[+] Direct imagining surveys can be used to better understand giant planet formation and put constraints on the entropies and luminosities of young giant planets. 
\item[+] Solar System missions: Within the solar system, various missions are dedicated to better understand planets and small bodies. Particularly relevant will be the missions to Uranus and/or Neptune \cite{2019arXiv190702963F}. The properties of these planets are poorly known, but it is now clear that they represent a category of planets that are numerous around other stars. 
\end{itemize}
\section{Challenges}  
As some questions are being solved, new questions arise; models reach the next level of complexity where more detailed processes must be included; different pieces of the puzzle must be fit into a unified model. We reach a point where the details matter, which makes models more complex but also allow achieving a deeper  understanding of the topic. 


We envision future progress in the following directions: 
\begin{itemize}
\item[+]  {\bf Combined solar-system and exoplanetary science:}
While solar system science focuses on explaining the detailed properties of the objects in our specific planetary system, exoplanetary science provides a broader view of planetary systems allowing to {\bf build} statistics and identify trends and correlations. However, each exoplanetary system is much less observationally constrained than our own. These two fields are clearly complementary and we urge  to bridge the two in future research. 

\item[+]  {\bf Combining dynamics with thermodynamics:} 
At the moment planet formation models can be broadly divided in two categories: some are based on a careful treatment of the thermodynamics and other physical processes, but assume that the planets form {\it in situ} and independently from the other objects of the system; others are based on dynamical simulations that model properly the interaction of the growing planet with the disk and the other objects, but the physical structure of the planet is modeled in an over-simplified way. 
Clearly, a unified formation model is required in the future. 

\item[+]  {\bf Linking current-state structure and origin:}

In order to link the internal structure of a planet with its origin, a good understanding of its evolution is required. 
This is because the three aspects of formation, evolution and internal structure are linked: the formation process determines the primordial planetary composition, internal structure and thermal state.  
These determine the heat transport mechanism, the potential re-distribution of the materials, and the planetary long-term evolution which could also include evaporation and outgassing. 
The planetary evolution then determines the current-state internal structure. As a result, in order to use the current-state planetary properties with formation, modeling the planetary evolution properly is required. 



\end{itemize}

The challenges in planet formation theory should not be viewed as obstacles but as exciting opportunities to advance the field and reach a new level of understanding of how planets form. After all, we live on a planet which is part of a planetary system, and only with a better knowledge of how planets form can we understand our uniqueness. 


\newpage


\begin{thebibliography}{}

\bibitem{2014prpl.conf..475A}
R.~{Alexander}, I.~{Pascucci}, S.~{Andrews}, P.~{Armitage}, and L.~{Cieza}.
\newblock {The Dispersal of Protoplanetary Disks}.
\newblock In Henrik {Beuther}, Ralf~S. {Klessen}, Cornelis~P. {Dullemond}, and
  Thomas {Henning}, editors, {\em Protostars and Planets VI}, page 475, January
  2014.

\bibitem{Alibert2005}
Yann Alibert, Christoph Mordasini, Willy Benz, and Christophe Winisdoerffer.
\newblock Models of giant planet formation with migration and disc evolution.
\newblock {\em Astronomy \& Astrophysics}, 434(1):343--353, 2005.

\bibitem{Alibert2018}
Yann Alibert, Julia Venturini, Ravit Helled, Sareh Ataiee, Remo Burn, Luc
  Senecal, Willy Benz, Lucio Mayer, Christoph Mordasini, Sascha~P Quanz, et~al.
\newblock The formation of jupiter by hybrid pebble--planetesimal accretion.
\newblock {\em Nature astronomy}, 2(11):873--877, 2018.

\bibitem{2020arXiv200105007A}
Sean~M. {Andrews}.
\newblock {Observations of Protoplanetary Disk Structures}.
\newblock {\em arXiv e-prints}, page arXiv:2001.05007, January 2020.

\bibitem{2016ApJ...828...46A}
M.~{Ansdell}, J.~P. {Williams}, and N.~et~al. {van der Marel}.
\newblock {ALMA Survey of Lupus Protoplanetary Disks. I. Dust and Gas Masses}.
\newblock {\em ApJ}, 828(1):46, September 2016.

\bibitem{2016ApJ...829..114B}
Konstantin {Batygin}, Peter~H. {Bodenheimer}, and Gregory~P. {Laughlin}.
\newblock {In Situ Formation and Dynamical Evolution of Hot Jupiter Systems}.
\newblock {\em ApJ}, 829(2):114, October 2016.

\bibitem{2012ApJ...751..119B}
C.~{Beaug{\'e}} and D.~{Nesvorn{\'y}}.
\newblock {Multiple-planet Scattering and the Origin of Hot Jupiters}.
\newblock {\em ApJ}, 751(2):119, June 2012.

\bibitem{2014prpl.conf..691B}
W.~{Benz}, S.~{Ida}, Y.~{Alibert}, D.~{Lin}, and C.~{Mordasini}.
\newblock {Planet Population Synthesis}.
\newblock In Henrik {Beuther}, Ralf~S. {Klessen}, Cornelis~P. {Dullemond}, and
  Thomas {Henning}, editors, {\em Protostars and Planets VI}, page 691, January
  2014.

\bibitem{2019A&A...623A..88B}
Bertram {Bitsch}, Andre {Izidoro}, Anders {Johansen}, Sean~N. {Raymond},
  Alessandro {Morbidelli}, Michiel {Lambrechts}, and Seth~A. {Jacobson}.
\newblock {Formation of planetary systems by pebble accretion and migration:
  growth of gas giants}.
\newblock {\em AAP}, 623:A88, March 2019.

\bibitem{2015A&A...582A.112B}
Bertram {Bitsch}, Michiel {Lambrechts}, and Anders {Johansen}.
\newblock {The growth of planets by pebble accretion in evolving protoplanetary
  discs}.
\newblock {\em AAP}, 582:A112, October 2015.

\bibitem{Bodenheimer+86}
P.~{Bodenheimer} and J.~B. {Pollack}.
\newblock {Calculations of the accretion and evolution of giant planets: The
  effects of solid cores}.
\newblock {\em Icarus}, 67(3):391--408, Sep 1986.

\bibitem{2009ApJ...695L..53B}
Aaron~C. {Boley}.
\newblock {The Two Modes of Gas Giant Planet Formation}.
\newblock {\em ApJL}, 695(1):L53--L57, April 2009.

\bibitem{2019NatAs...3..416B}
Aldo~S. {Bonomo}, Li~{Zeng}, and Mario et~al. {Damasso}.
\newblock {A giant impact as the likely origin of different twins in the
  Kepler-107 exoplanet system}.
\newblock {\em Nature Astronomy}, 3:416--423, February 2019.

\bibitem{Boss1997}
Alan~P Boss.
\newblock Giant planet formation by gravitational instability.
\newblock {\em Science}, 276(5320):1836--1839, 1997.

\bibitem{2013EPJWC..4703005B}
C.~{Broeg}, A.~{Fortier}, and D.~et~al. {Ehrenreich}.
\newblock {CHEOPS: A transit photometry mission for ESA's small mission
  programme}.
\newblock In {\em European Physical Journal Web of Conferences}, volume~47 of
  {\em European Physical Journal Web of Conferences}, page 03005, April 2013.

\bibitem{Brouwers20}
{Brouwers} and {Ormel}.
\newblock How planets grow by pebble accretion - ii. analytical calculations on
  the evolution of polluted envelopes.
\newblock {\em A\&A}, 634:A15, 2020.

\bibitem{2020arXiv200604121B}
Natacha {Br{\"u}gger}, Remo {Burn}, Gavin {Coleman}, Yann {Alibert}, and Willy
  {Benz}.
\newblock {Pebbles versus planetesimals: the outcomes of population synthesis
  models}.
\newblock {\em arXiv e-prints}, page arXiv:2006.04121, June 2020.

\bibitem{2018SciA....4.6887C}
Robin {Canup} and Julien {Salmon}.
\newblock {Origin of Phobos and Deimos by the impact of a Vesta-to-Ceres sized
  body with Mars}.
\newblock {\em Science Advances}, 4(4):eaar6887, April 2018.

\bibitem{2005Sci...307..546C}
Robin~M. {Canup}.
\newblock {A Giant Impact Origin of Pluto-Charon}.
\newblock {\em Science}, 307(5709):546--550, January 2005.

\bibitem{2001Natur.412..708C}
Robin~M. {Canup} and Erik {Asphaug}.
\newblock {Origin of the Moon in a giant impact near the end of the Earth's
  formation}.
\newblock {\em Nature}, 412(6848):708--712, August 2001.

\bibitem{2014ApJ...780...53C}
Sourav {Chatterjee} and Jonathan~C. {Tan}.
\newblock {Inside-out Planet Formation}.
\newblock {\em ApJ}, 780(1):53, January 2014.

\bibitem{2018ApJ...865...35C}
Alice {Chau}, Christian {Reinhardt}, Ravit {Helled}, and Joachim {Stadel}.
\newblock {Forming Mercury by Giant Impacts}.
\newblock {\em ApJ}, 865(1):35, September 2018.

\bibitem{2008PASP..120..531C}
Andrew {Cumming}, R.~Paul {Butler}, Geoffrey~W. {Marcy}, Steven~S. {Vogt},
  Jason~T. {Wright}, and Debra~A. {Fischer}.
\newblock {The Keck Planet Search: Detectability and the Minimum Mass and
  Orbital Period Distribution of Extrasolar Planets}.
\newblock {\em PASP}, 120(867):531, May 2008.

\bibitem{2008ApJ...687.1432C}
Jeffrey~N. {Cuzzi}, Robert~C. {Hogan}, and Karim {Shariff}.
\newblock {Toward Planetesimals: Dense Chondrule Clumps in the Protoplanetary
  Nebula}.
\newblock {\em ApJ}, 687(2):1432--1447, November 2008.

\bibitem{2010exop.book..319D}
G.~{D'Angelo}, R.~H. {Durisen}, and J.~J. {Lissauer}.
\newblock {\em {Giant Planet Formation}}, pages 319--346.
\newblock 2010.

\bibitem{2003ApJ...586..540D}
Gennaro {D'Angelo}, Willy {Kley}, and Thomas {Henning}.
\newblock {Orbital Migration and Mass Accretion of Protoplanets in
  Three-dimensional Global Computations with Nested Grids}.
\newblock {\em ApJ}, 586(1):540--561, March 2003.

\bibitem{2018haex.bookE.140D}
Gennaro {D'Angelo} and Jack~J. {Lissauer}.
\newblock {\em {Formation of Giant Planets}}, page 140.
\newblock 2018.

\bibitem{2011Natur.473..489D}
N.~{Dauphas} and A.~{Pourmand}.
\newblock {Hf-W-Th evidence for rapid growth of Mars and its status as a
  planetary embryo}.
\newblock {\em Nature}, 473(7348):489--492, May 2011.

\bibitem{Debras19}
Florian Debras and Gilles Chabrier.
\newblock {\em The Astrophysical Journal}, 872(1):100, feb 2019.

\bibitem{2017A&A...608A..92D}
J.~{Drazkowska} and Y.~{Alibert}.
\newblock {Planetesimal formation starts at the snow line}.
\newblock {\em AAP}, 608:A92, December 2017.

\bibitem{Durisen2007}
R.~H. {Durisen}, A.~P. {Boss}, L.~{Mayer}, A.~F. {Nelson}, T.~{Quinn}, and
  W.~K.~M. {Rice}.
\newblock {Gravitational Instabilities in Gaseous Protoplanetary Disks and
  Implications for Giant Planet Formation}.
\newblock In Bo~{Reipurth}, David {Jewitt}, and Klaus {Keil}, editors, {\em
  Protostars and Planets V}, page 607, Jan 2007.

\bibitem{2005ApJ...622.1102F}
Debra~A. {Fischer} and Jeff {Valenti}.
\newblock {The Planet-Metallicity Correlation}.
\newblock {\em ApJ}, 622(2):1102--1117, April 2005.

\bibitem{2019arXiv190702963F}
Leigh N. et~al. {Fletcher}.
\newblock {Ice Giant Systems: The Scientific Potential of Orbital Missions to
  Uranus and Neptune}.
\newblock {\em arXiv e-prints}, page arXiv:1907.02963, July 2019.

\bibitem{2019A&A...630A.147F}
Mario {Flock}, Neal~J. {Turner}, Gijs~D. {Mulders}, Yasuhiro {Hasegawa},
  Richard~P. {Nelson}, and Bertram {Bitsch}.
\newblock {Planet formation and migration near the silicate sublimation front
  in protoplanetary disks}.
\newblock {\em AAP}, 630:A147, October 2019.

\bibitem{2018MNRAS.474.5036F}
D.~H. {Forgan}, C.~{Hall}, F.~{Meru}, and W.~K.~M. {Rice}.
\newblock {Towards a population synthesis model of self-gravitating disc
  fragmentation and tidal downsizing II: the effect of fragment-fragment
  interactions}.
\newblock {\em MNRAS}, 474(4):5036--5048, March 2018.

\bibitem{1980ApJ...241..425G}
P.~{Goldreich} and S.~{Tremaine}.
\newblock {Disk-satellite interactions.}
\newblock {\em ApJ}, 241:425--441, October 1980.

\bibitem{2016ApJ...817...17G}
Thomas~P. {Greene}, Michael~R. {Line}, Cezar {Montero}, Jonathan~J. {Fortney},
  Jacob {Lustig-Yaeger}, and Kyle {Luther}.
\newblock {Characterizing Transiting Exoplanet Atmospheres with JWST}.
\newblock {\em ApJ}, 817(1):17, January 2016.

\bibitem{2010A&A...513A..56G}
C.~{G{\"u}ttler}, J.~{Blum}, A.~{Zsom}, C.~W. {Ormel}, and C.~P. {Dullemond}.
\newblock {The outcome of protoplanetary dust growth: pebbles, boulders, or
  planetesimals?. I. Mapping the zoo of laboratory collision experiments}.
\newblock {\em AAP}, 513:A56, April 2010.

\bibitem{2020ApJ...892..120H}
Thomas {Hartlep} and Jeffrey~N. {Cuzzi}.
\newblock {Cascade Model for Planetesimal Formation by Turbulent Clustering}.
\newblock {\em ApJ}, 892(2):120, April 2020.

\bibitem{HelledStevenson17}
Helled and Stevenson.
\newblock {\em The Astrophysical Journal}, 840(1):L4, apr 2017.

\bibitem{Helled2014}
R.~{Helled}, P.~{Bodenheimer}, M.~{Podolak}, A.~{Boley}, F.~{Meru},
  S.~{Nayakshin}, J.~J. {Fortney}, L.~{Mayer}, Y.~{Alibert}, and A.~P. {Boss}.
\newblock {Giant Planet Formation, Evolution, and Internal Structure}.
\newblock In Henrik {Beuther}, Ralf~S. {Klessen}, Cornelis~P. {Dullemond}, and
  Thomas {Henning}, editors, {\em Protostars and Planets VI}, page 643, Jan
  2014.

\bibitem{2014ApJ...789...69H}
Ravit {Helled} and Peter {Bodenheimer}.
\newblock {The Formation of Uranus and Neptune: Challenges and Implications for
  Intermediate-mass Exoplanets}.
\newblock {\em ApJ}, 789(1):69, July 2014.

\bibitem{2018haex.bookE..44H}
Ravit {Helled} and Tristan {Guillot}.
\newblock {\em {Internal Structure of Giant and Icy Planets: Importance of
  Heavy Elements and Mixing}}, page~44.
\newblock 2018.

\bibitem{2016A&A...596L...3I}
S.~{Ida} and T.~{Guillot}.
\newblock {Formation of dust-rich planetesimals from sublimated pebbles inside
  of the snow line}.
\newblock {\em AAP}, 596:L3, November 2016.

\bibitem{1993Icar..106..210I}
Shigeru {Ida} and Junichiro {Makino}.
\newblock {Scattering of Planetesimals by a Protoplanet: Slowing Down of
  Runaway Growth}.
\newblock {\em Icarus}, 106(1):210--227, November 1993.

\bibitem{2020arXiv200313582I}
Shigeru {Ida}, Shoji {Ueta}, Takanori {Sasaki}, and Yuya {Ishizawa}.
\newblock {Uranian Satellite Formation by Evolution of a Water Vapor Disk
  Generated by a Giant Impact}.
\newblock {\em arXiv e-prints}, page arXiv:2003.13582, March 2020.

\bibitem{2015MNRAS.448.1751I}
Niraj~K. {Inamdar} and Hilke~E. {Schlichting}.
\newblock {The formation of super-Earths and mini-Neptunes with giant impacts}.
\newblock {\em MNRAS}, 448(2):1751--1760, April 2015.

\bibitem{2018ApJ...853..163J}
Sheng {Jin} and Christoph {Mordasini}.
\newblock {Compositional Imprints in Density-Distance-Time: A Rocky Composition
  for Close-in Low-mass Exoplanets from the Location of the Valley of
  Evaporation}.
\newblock {\em ApJ}, 853(2):163, February 2018.

\bibitem{2006ApJ...643.1219J}
Anders {Johansen}, Thomas {Henning}, and Hubert {Klahr}.
\newblock {Dust Sedimentation and Self-sustained Kelvin-Helmholtz Turbulence in
  Protoplanetary Disk Midplanes}.
\newblock {\em ApJ}, 643(2):1219--1232, June 2006.

\bibitem{2007Natur.448.1022J}
Anders {Johansen}, Jeffrey~S. {Oishi}, Mordecai-Mark {Mac Low}, Hubert {Klahr},
  Thomas {Henning}, and Andrew {Youdin}.
\newblock {Rapid planetesimal formation in turbulent circumstellar disks}.
\newblock {\em Nature}, 448(7157):1022--1025, August 2007.

\bibitem{2009ApJ...704L..75J}
Anders {Johansen}, Andrew {Youdin}, and Mordecai-Mark {Mac Low}.
\newblock {Particle Clumping and Planetesimal Formation Depend Strongly on
  Metallicity}.
\newblock {\em ApJL}, 704(2):L75--L79, October 2009.

\bibitem{2018ApJ...861...52K}
J.~A. {Kegerreis}, L.~F.~A. {Teodoro}, V.~R. {Eke}, R.~J. {Massey}, D.~C.
  {Catling}, C.~L. {Fryer}, D.~G. {Korycansky}, M.~S. {Warren}, and K.~J.
  {Zahnle}.
\newblock {Consequences of Giant Impacts on Early Uranus for Rotation, Internal
  Structure, Debris, and Atmospheric Erosion}.
\newblock {\em ApJ}, 861(1):52, July 2018.

\bibitem{2018A&A...617A..44K}
M.~{Keppler}, M.~{Benisty}, A.~{M{\"u}ller}, and {Henning} et~al.
\newblock {Discovery of a planetary-mass companion within the gap of the
  transition disk around PDS 70}.
\newblock {\em AAP}, 617:A44, September 2018.

\bibitem{2009GeCoA..73.5150K}
Thorsten {Kleine}, Mathieu {Touboul}, Bernard {Bourdon}, Francis {Nimmo}, Klaus
  {Mezger}, Herbert {Palme}, Stein~B. {Jacobsen}, Qing-Zhu {Yin}, and
  Alexander~N. {Halliday}.
\newblock {Hf-W chronology of the accretion and early evolution of asteroids
  and terrestrial planets}.
\newblock {\em GCA}, 73(17):5150--5188, September 2009.

\bibitem{2016ARA&A..54..271K}
Kaitlin {Kratter} and Giuseppe {Lodato}.
\newblock {Gravitational Instabilities in Circumstellar Disks}.
\newblock {\em ARAA}, 54:271--311, September 2016.

\bibitem{2017PNAS..114.6712K}
Thomas~S. {Kruijer}, Christoph {Burkhardt}, Gerrit {Budde}, and Thorsten
  {Kleine}.
\newblock {Age of Jupiter inferred from the distinct genetics and formation
  times of meteorites}.
\newblock {\em Proceedings of the National Academy of Science},
  114(26):6712--6716, June 2017.

\bibitem{2010Sci...329...57L}
A.~M. {Lagrange}, M.~{Bonnefoy}, G.~{Chauvin}, D.~{Apai}, D.~{Ehrenreich},
  A.~{Boccaletti}, D.~{Gratadour}, D.~{Rouan}, D.~{Mouillet}, S.~{Lacour}, and
  M.~{Kasper}.
\newblock {A Giant Planet Imaged in the Disk of the Young Star
  {\ensuremath{\beta}} Pictoris}.
\newblock {\em Science}, 329(5987):57, July 2010.

\bibitem{2019A&A...630A..82L}
M.~{Lambrechts}, E.~{Lega}, R.~P. {Nelson}, A.~{Crida}, and A.~{Morbidelli}.
\newblock {Quasi-static contraction during runaway gas accretion onto giant
  planets}.
\newblock {\em AAP}, 630:A82, October 2019.

\bibitem{Lambrechts2012}
Michiel Lambrechts and Anders Johansen.
\newblock Rapid growth of gas-giant cores by pebble accretion.
\newblock {\em Astronomy \& Astrophysics}, 544:A32, 2012.

\bibitem{2019A&A...627A..83L}
Michiel {Lambrechts}, Alessandro {Morbidelli}, Seth~A. {Jacobson}, Anders
  {Johansen}, Bertram {Bitsch}, Andre {Izidoro}, and Sean~N. {Raymond}.
\newblock {Formation of planetary systems by pebble accretion and migration.
  How the radial pebble flux determines a terrestrial-planet or super-Earth
  growth mode}.
\newblock {\em AAP}, 627:A83, July 2019.

\bibitem{2020Icar..33913551L}
H.~{Lammer}, M.~{Leitzinger}, M.~{Scherf}, P.~{Odert}, C.~{Burger},
  D.~{Kubyshkina}, C.~{Johnstone}, T.~{Maindl}, C.~M. {Sch{\"a}fer},
  M.~{G{\"u}del}, N.~{Tosi}, A.~{Nikolaou}, E.~{Marcq}, N.~V. {Erkaev},
  L.~{Noack}, K.~G. {Kislyakova}, L.~{Fossati}, E.~{Pilat-Lohinger},
  F.~{Ragossnig}, and E.~A. {Dorfi}.
\newblock {Constraining the early evolution of Venus and Earth through
  atmospheric Ar, Ne isotope and bulk K/U ratios}.
\newblock {\em Icarus}, 339:113551, March 2020.

\bibitem{2012A&A...540A..20L}
J.~{Leconte} and G.~{Chabrier}.
\newblock {A new vision of giant planet interiors: Impact of double diffusive
  convection}.
\newblock {\em AAP}, 540:A20, April 2012.

\bibitem{2010AJ....139.1297L}
Harold~F. {Levison}, Edward {Thommes}, and Martin~J. {Duncan}.
\newblock {Modeling the Formation of Giant Planet Cores. I. Evaluating Key
  Processes}.
\newblock {\em AJ}, 139(4):1297--1314, April 2010.

\bibitem{2009Icar..199..338L}
Jack~J. {Lissauer}, Olenka {Hubickyj}, Gennaro {D'Angelo}, and Peter
  {Bodenheimer}.
\newblock {Models of Jupiter's growth incorporating thermal and hydrodynamic
  constraints}.
\newblock {\em Icarus}, 199(2):338--350, February 2009.

\bibitem{2015MNRAS.446.1685L}
Shang-Fei {Liu}, Craig~B. {Agnor}, D.~N.~C. {Lin}, and Shu-Lin {Li}.
\newblock {Embryo impacts and gas giant mergers - II. Diversity of hot
  Jupiters' internal structure}.
\newblock {\em MNRAS}, 446(2):1685--1702, January 2015.

\bibitem{Shang19}
Shang-Fei Liu, Yasunori Hori, Simon Müller, Xiaochen Zheng, Ravit Helled, Doug
  Lin, and Andrea Isella.
\newblock The formation of jupiter’s diluted core by a giant impact.
\newblock {\em Nature}, 572:355--357, 08 2019.

\bibitem{2018ApJ...866...49L}
M.~{Lozovsky}, R.~{Helled}, C.~{Dorn}, and J.~{Venturini}.
\newblock {Threshold Radii of Volatile-rich Planets}.
\newblock {\em ApJ}, 866(1):49, October 2018.

\bibitem{2017ApJ...836..227L}
Michael {Lozovsky}, Ravit {Helled}, Eric~D. {Rosenberg}, and Peter
  {Bodenheimer}.
\newblock {Jupiter{\textquoteright}s Formation and Its Primordial Internal
  Structure}.
\newblock {\em ApJ}, 836(2):227, February 2017.

\bibitem{1999ApJ...526.1001L}
S.~H. {Lubow}, M.~{Seibert}, and P.~{Artymowicz}.
\newblock {Disk Accretion onto High-Mass Planets}.
\newblock {\em ApJ}, 526(2):1001--1012, December 1999.

\bibitem{2009AIPC.1158....3M}
Eric~E. {Mamajek}.
\newblock {Initial Conditions of Planet Formation: Lifetimes of Primordial
  Disks}.
\newblock In Tomonori {Usuda}, Motohide {Tamura}, and Miki {Ishii}, editors,
  {\em American Institute of Physics Conference Series}, volume 1158 of {\em
  American Institute of Physics Conference Series}, pages 3--10, August 2009.

\bibitem{2008Natur.453.1216M}
Margarita~M. {Marinova}, Oded {Aharonson}, and Erik {Asphaug}.
\newblock {Mega-impact formation of the Mars hemispheric dichotomy}.
\newblock {\em Nature}, 453(7199):1216--1219, June 2008.

\bibitem{2010Natur.468.1080M}
Christian {Marois}, B.~{Zuckerman}, Quinn~M. {Konopacky}, Bruce {Macintosh},
  and Travis {Barman}.
\newblock {Images of a fourth planet orbiting HR 8799}.
\newblock {\em Nature}, 468(7327):1080--1083, December 2010.

\bibitem{2017Sci...356.1069M}
B.~{Marty}, K.~{Altwegg}, H.~{Balsiger}, A.~{Bar-Nun}, D.~V. {Bekaert}, J.~J.
  {Berthelier}, A.~{Bieler}, C.~{Briois}, U.~{Calmonte}, M.~{Combi}, J.~{De
  Keyser}, B.~{Fiethe}, S.~A. {Fuselier}, S.~{Gasc}, T.~I. {Gombosi}, K.~C.
  {Hansen}, M.~{H{\"a}ssig}, A.~{J{\"a}ckel}, E.~{Kopp}, A.~{Korth}, L.~{Le
  Roy}, U.~{Mall}, O.~{Mousis}, T.~{Owen}, H.~{R{\`e}me}, M.~{Rubin},
  T.~{S{\'e}mon}, C.~Y. {Tzou}, J.~H. {Waite}, and P.~{Wurz}.
\newblock {Xenon isotopes in 67P/Churyumov-Gerasimenko show that comets
  contributed to Earth's atmosphere}.
\newblock {\em Science}, 356(6342):1069--1072, June 2017.

\bibitem{2016E&PSL.441...91M}
Bernard {Marty}, Guillaume {Avice}, Yuji {Sano}, Kathrin {Altwegg}, Hans
  {Balsiger}, Myrtha {H{\"a}ssig}, Alessandro {Morbidelli}, Olivier {Mousis},
  and Martin {Rubin}.
\newblock {Origins of volatile elements (H, C, N, noble gases) on Earth and
  Mars in light of recent results from the ROSETTA cometary mission}.
\newblock {\em Earth and Planetary Science Letters}, 441:91--102, May 2016.

\bibitem{2001MNRAS.320L..55M}
F.~{Masset} and M.~{Snellgrove}.
\newblock {Reversing type II migration: resonance trapping of a lighter giant
  protoplanet}.
\newblock {\em MNRAS}, 320(4):L55--L59, February 2001.

\bibitem{2014prpl.conf..521M}
B.~C. {Matthews}, A.~V. {Krivov}, M.~C. {Wyatt}, G.~{Bryden}, and C.~{Eiroa}.
\newblock {Observations, Modeling, and Theory of Debris Disks}.
\newblock In Henrik {Beuther}, Ralf~S. {Klessen}, Cornelis~P. {Dullemond}, and
  Thomas {Henning}, editors, {\em Protostars and Planets VI}, page 521, January
  2014.

\bibitem{Mayer2002}
Lucio Mayer, Thomas Quinn, James Wadsley, and Joachim Stadel.
\newblock Formation of giant planets by fragmentation of protoplanetary disks.
\newblock {\em Science}, 298(5599):1756--1759, 2002.

\bibitem{2011arXiv1109.2497M}
M.~{Mayor}, M.~{Marmier}, C.~{Lovis}, S.~{Udry}, D.~{S{\'e}gransan}, F.~{Pepe},
  W.~{Benz}, J.~L. {Bertaux}, F.~{Bouchy}, X.~{Dumusque}, G.~{Lo Curto},
  C.~{Mordasini}, D.~{Queloz}, and N.~C. {Santos}.
\newblock {The HARPS search for southern extra-solar planets XXXIV. Occurrence,
  mass distribution and orbital properties of super-Earths and Neptune-mass
  planets}.
\newblock {\em arXiv e-prints}, page arXiv:1109.2497, September 2011.

\bibitem{mayor_queloz_1995}
Michel {Mayor} and Didier {Queloz}.
\newblock {A Jupiter-mass companion to a solar-type star}.
\newblock {\em Nature}, 378(6555):355--359, Nov 1995.

\bibitem{2011ApJ...736L..29M}
Neil {Miller} and Jonathan~J. {Fortney}.
\newblock {The Heavy-element Masses of Extrasolar Giant Planets, Revealed}.
\newblock {\em ApJL}, 736(2):L29, August 2011.

\bibitem{2019Sci...365.1441M}
J.~C. {Morales}, A.~J. {Mustill}, I.~{Ribas}, and et~al. {Davies}.
\newblock {A giant exoplanet orbiting a very-low-mass star challenges planet
  formation models}.
\newblock {\em Science}, 365(6460):1441--1445, September 2019.

\bibitem{2014Icar..232..266M}
A.~{Morbidelli}, J.~{Szul{\'a}gyi}, A.~{Crida}, E.~{Lega}, B.~{Bitsch},
  T.~{Tanigawa}, and K.~{Kanagawa}.
\newblock {Meridional circulation of gas into gaps opened by giant planets in
  three-dimensional low-viscosity disks}.
\newblock {\em Icarus}, 232:266--270, April 2014.

\bibitem{2007AJ....134.1790M}
Alessandro {Morbidelli}, Kleomenis {Tsiganis}, Aur{\'e}lien {Crida}, Harold~F.
  {Levison}, and Rodney {Gomes}.
\newblock {Dynamics of the Giant Planets of the Solar System in the Gaseous
  Protoplanetary Disk and Their Relationship to the Current Orbital
  Architecture}.
\newblock {\em AJ}, 134(5):1790--1798, November 2007.

\bibitem{2018haex.bookE.143M}
Christoph {Mordasini}.
\newblock {\em {Planetary Population Synthesis}}, page 143.
\newblock 2018.

\bibitem{2010Icar..209..616M}
Naor {Movshovitz}, Peter {Bodenheimer}, Morris {Podolak}, and Jack~J.
  {Lissauer}.
\newblock {Formation of Jupiter using opacities based on detailed grain
  physics}.
\newblock {\em Icarus}, 209(2):616--624, October 2010.

\bibitem{2020arXiv200413534M}
Simon {M{\"u}ller}, Ravit {Helled}, and Andrew {Cumming}.
\newblock {The Challenge of Forming a Fuzzy Core in Jupiter}.
\newblock {\em arXiv e-prints}, page arXiv:2004.13534, April 2020.

\bibitem{2018ApJ...854..112M}
Simon {M{\"u}ller}, Ravit {Helled}, and Lucio {Mayer}.
\newblock {On the Diversity in Mass and Orbital Radius of Giant Planets Formed
  via Disk Instability}.
\newblock {\em ApJ}, 854(2):112, February 2018.

\bibitem{Nayakshin2017}
Sergei {Nayakshin}.
\newblock {Dawes Review 7: The Tidal Downsizing Hypothesis of Planet
  Formation}.
\newblock {\em Publications of the Astronomical Society of Australia}, 34:e002,
  January 2017.

\bibitem{2018ARA&A..56..137N}
David {Nesvorn{\'y}}.
\newblock {Dynamical Evolution of the Early Solar System}.
\newblock {\em ARAA}, 56:137--174, September 2018.

\bibitem{2020A&A...634A..43O}
J.~F. {Otegi}, F.~{Bouchy}, and R.~{Helled}.
\newblock {Revisited mass-radius relations for exoplanets below 120
  M$_{{\ensuremath{\oplus}}}$}.
\newblock {\em AAP}, 634:A43, February 2020.

\bibitem{2017ApJ...847...29O}
James~E. {Owen} and Yanqin {Wu}.
\newblock {The Evaporation Valley in the Kepler Planets}.
\newblock {\em ApJ}, 847(1):29, September 2017.

\bibitem{2018SSRv..214...38P}
Sijme-Jan {Paardekooper} and Anders {Johansen}.
\newblock {Giant Planet Formation and Migration}.
\newblock {\em SSR}, 214(1):38, February 2018.

\bibitem{Pollack1996}
James~B Pollack, Olenka Hubickyj, Peter Bodenheimer, Jack~J Lissauer, Morris
  Podolak, and Yuval Greenzweig.
\newblock Formation of the giant planets by concurrent accretion of solids and
  gas.
\newblock {\em Icarus}, 124(1):62--85, 1996.

\bibitem{2018MNRAS.479.5136P}
Andrius {Popovas}, {\r{A}}ke {Nordlund}, Jon~P. {Ramsey}, and Chris~W. {Ormel}.
\newblock {Pebble dynamics and accretion on to rocky planets - I. Adiabatic and
  convective models}.
\newblock {\em MNRAS}, 479(4):5136--5156, October 2018.

\bibitem{2014ExA....38..249R}
H.~{Rauer}, C.~{Catala}, and {Aerts} et~al.
\newblock {The PLATO 2.0 mission}.
\newblock {\em Experimental Astronomy}, 38(1-2):249--330, November 2014.

\bibitem{2020MNRAS.492.5336R}
Christian {Reinhardt}, Alice {Chau}, Joachim {Stadel}, and Ravit {Helled}.
\newblock {Bifurcation in the history of Uranus and Neptune: the role of giant
  impacts}.
\newblock {\em MNRAS}, 492(4):5336--5353, March 2020.

\bibitem{1969edo..book.....S}
Viktor~Sergeevich {Safronov}.
\newblock {\em {Evoliutsiia doplanetnogo oblaka.}}
\newblock 1969.

\bibitem{2017A&A...603A..30S}
N.~C. {Santos}, V.~{Adibekyan}, P.~{Figueira}, D.~T. {Andreasen}, S.~C.~C.
  {Barros}, E.~{Delgado-Mena}, O.~{Demangeon}, J.~P. {Faria}, M.~{Oshagh},
  S.~G. {Sousa}, P.~T.~P. {Viana}, and A.~C.~S. {Ferreira}.
\newblock {Observational evidence for two distinct giant planet populations}.
\newblock {\em AAP}, 603:A30, July 2017.

\bibitem{2018Natur.555..507S}
Martin {Schiller}, Martin {Bizzarro}, and Vera~Assis {Fernandes}.
\newblock {Isotopic evolution of the protoplanetary disk and the building
  blocks of Earth and the Moon}.
\newblock {\em Nature}, 555(7697):507--510, March 2018.

\bibitem{2017A&A...602A..21S}
Djoeke {Schoonenberg} and Chris~W. {Ormel}.
\newblock {Planetesimal formation near the snowline: in or out?}
\newblock {\em AAP}, 602:A21, June 2017.

\bibitem{Shibata+19b}
Sho {Shibata}, Ravit {Helled}, and Masahiro {Ikoma}.
\newblock {The origin of the high metallicity of close-in giant exoplanets:
  Combined effect of the resonant and aerodynamic shepherding}.
\newblock {\em arXiv e-prints}, page arXiv:1911.02292, Nov 2019.

\bibitem{Shibata+19a}
Sho {Shibata} and Masahiro {Ikoma}.
\newblock {Capture of solids by growing proto-gas giants: effects of gap
  formation and supply limited growth}.
\newblock {\em MNRAS}, 487(4):4510--4524, Aug 2019.

\bibitem{1982P&SS...30..755S}
D.~J. {Stevenson}.
\newblock {Formation of the giant planets}.
\newblock {\em PLANSS}, 30(8):755--764, August 1982.

\bibitem{2020ApJ...891..143T}
Hidekazu {Tanaka}, Kiyoka {Murase}, and Takayuki {Tanigawa}.
\newblock {Final Masses of Giant Planets. III. Effect of Photoevaporation and a
  New Planetary Migration Model}.
\newblock {\em ApJ}, 891(2):143, March 2020.

\bibitem{2002ApJ...565.1257T}
Hidekazu {Tanaka}, Taku {Takeuchi}, and William~R. {Ward}.
\newblock {Three-Dimensional Interaction between a Planet and an Isothermal
  Gaseous Disk. I. Corotation and Lindblad Torques and Planet Migration}.
\newblock {\em ApJ}, 565(2):1257--1274, February 2002.

\bibitem{2007ApJ...667..557T}
Takayuki {Tanigawa} and Masahiro {Ikoma}.
\newblock {A Systematic Study of the Final Masses of Gas Giant Planets}.
\newblock {\em ApJ}, 667(1):557--570, September 2007.

\bibitem{2016ApJ...823...48T}
Takayuki {Tanigawa} and Hidekazu {Tanaka}.
\newblock {Final Masses of Giant Planets. II. Jupiter Formation in a
  Gas-depleted Disk}.
\newblock {\em ApJ}, 823(1):48, May 2016.

\bibitem{2014prpl.conf..339T}
L.~{Testi}, T.~{Birnstiel}, L.~{Ricci}, S.~{Andrews}, J.~{Blum},
  J.~{Carpenter}, C.~{Dominik}, A.~{Isella}, A.~{Natta}, J.~P. {Williams}, and
  D.~J. {Wilner}.
\newblock {Dust Evolution in Protoplanetary Disks}.
\newblock In Henrik {Beuther}, Ralf~S. {Klessen}, Cornelis~P. {Dullemond}, and
  Thomas {Henning}, editors, {\em Protostars and Planets VI}, page 339, January
  2014.

\bibitem{2019RNAAS...3..128T}
Daniel~P. {Thorngren}, Mark~S. {Marley}, and Jonathan~J. {Fortney}.
\newblock {An Empirical Mass-Radius Relation for Cool Giant Planets}.
\newblock {\em Research Notes of the American Astronomical Society}, 3(9):128,
  September 2019.

\bibitem{2018ExA....46..135T}
Giovanna {Tinetti}, Pierre {Drossart}, Paul {Eccleston}, and et~al. {Hartogh},
  Paul.
\newblock {A chemical survey of exoplanets with ARIEL}.
\newblock {\em Experimental Astronomy}, 46(1):135--209, November 2018.

\bibitem{1964ApJ...139.1217T}
A.~{Toomre}.
\newblock {On the gravitational stability of a disk of stars.}
\newblock {\em ApJ}, 139:1217--1238, May 1964.

\bibitem{valleta18}
Claudio {Valletta} and Ravit {Helled}.
\newblock {The distribution of heavy-elements in giant protoplanetary
  atmospheres: the importance of planetesimal-envelope interactions}.
\newblock {\em arXiv e-prints}, page arXiv:1811.10904, Nov 2018.

\bibitem{2018A&A...610L..14V}
Allona {Vazan}, Ravit {Helled}, and Tristan {Guillot}.
\newblock {Jupiter's evolution with primordial composition gradients}.
\newblock {\em AAP}, 610:L14, February 2018.

\bibitem{2016A&A...596A..90V}
Julia {Venturini}, Yann {Alibert}, and Willy {Benz}.
\newblock {Planet formation with envelope enrichment: new insights on planetary
  diversity}.
\newblock {\em AAP}, 596:A90, December 2016.

\bibitem{2017ApJ...848...95V}
Julia {Venturini} and Ravit {Helled}.
\newblock {The Formation of Mini-Neptunes}.
\newblock {\em ApJ}, 848(2):95, October 2017.

\bibitem{2020A&A...634A..31V}
Julia {Venturini} and Ravit {Helled}.
\newblock {Jupiter's heavy-element enrichment expected from formation models}.
\newblock {\em AAP}, 634:A31, February 2020.

\bibitem{wahl2017}
S.~M. Wahl, W.~B. Hubbard, B.~Militzer, T.~Guillot, Y.~Miguel, N.~Movshovitz,
  Y.~Kaspi, R.~Helled, D.~Reese, E.~Galanti, S.~Levin, J.~E. Connerney, and
  S.~J. Bolton.
\newblock {\em Geophysical Research Letters}, 44(10):4649--4659, 2017.

\bibitem{2011Natur.475..206W}
Kevin~J. {Walsh}, Alessandro {Morbidelli}, Sean~N. {Raymond}, David~P.
  {O'Brien}, and Avi~M. {Mandell}.
\newblock {A low mass for Mars from Jupiter's early gas-driven migration}.
\newblock {\em Nature}, 475(7355):206--209, July 2011.

\bibitem{1992Icar..100..307W}
George~W. {Wetherill}.
\newblock {An alternative model for the formation of the asteroids}.
\newblock {\em Icarus}, 100(2):307--325, December 1992.

\bibitem{2011ARA&A..49...67W}
Jonathan~P. {Williams} and Lucas~A. {Cieza}.
\newblock {Protoplanetary Disks and Their Evolution}.
\newblock {\em ARAA}, 49(1):67--117, September 2011.

\bibitem{2000prpl.conf.1081W}
G.~{Wuchterl}, T.~{Guillot}, and J.~J. {Lissauer}.
\newblock {Giant Planet Formation}.
\newblock In V.~{Mannings}, A.~P. {Boss}, and S.~S. {Russell}, editors, {\em
  Protostars and Planets IV}, page 1081, May 2000.

\bibitem{2017A&A...606A..80Y}
C.~C. {Yang}, A.~{Johansen}, and D.~{Carrera}.
\newblock {Concentrating small particles in protoplanetary disks through the
  streaming instability}.
\newblock {\em AAP}, 606:A80, October 2017.

\bibitem{2005ApJ...620..459Y}
Andrew~N. {Youdin} and Jeremy {Goodman}.
\newblock {Streaming Instabilities in Protoplanetary Disks}.
\newblock {\em ApJ}, 620(1):459--469, February 2005.

\bibitem{2010A&A...513A..57Z}
A.~{Zsom}, C.~W. {Ormel}, C.~{G{\"u}ttler}, J.~{Blum}, and C.~P. {Dullemond}.
\newblock {The outcome of protoplanetary dust growth: pebbles, boulders, or
  planetesimals? II. Introducing the bouncing barrier}.
\newblock {\em AAP}, 513:A57, April 2010.

\end{thebibliography}
\end{document}